%% file: aspconArXiv.tex
\documentclass[conference]{IEEEtran}
\IEEEoverridecommandlockouts
\usepackage{cite}
\usepackage{amsmath,amssymb,amsfonts}
\usepackage{algorithmic}
\usepackage{graphicx}
\usepackage{textcomp}
\usepackage{xcolor}
\def\BibTeX{{\rm B\kern-.05em{\sc i\kern-.025em b}\kern-.08em
    T\kern-.1667em\lower.7ex\hbox{E}\kern-.125emX}}

\usepackage{amsthm}

\usepackage{tikz}
\usetikzlibrary{arrows}
\usepackage{pgfplots}

\newcommand{\av}{\ensuremath{\mathbf{a}}}

\newcommand{\Jm}{\ensuremath{\mathbf{J}}}
\newcommand{\Nm}{\ensuremath{\mathbf{N}}}

\newcommand{\uv}{\ensuremath{\mathbf{u}}}
\newcommand{\Vm}{\ensuremath{\mathbf{V}}}
\newcommand{\vv}{\ensuremath{\mathbf{v}}}

\newcommand{\zerov}{\ensuremath{\mathbf{0}}}

\graphicspath{{FiguresNMI/}}

\pgfplotsset{compat=1.14} 

\begin{document}

\title{Enhanced Normalized Mutual Information\\
for Localization in Noisy Environments
\thanks{
This material is based upon work supported, in part, by Ford Autonomous Vehicles LLC.
Portions of this research were conducted with the advanced computing resources provided by Texas A\&M High Performance Research Computing.}}

\author{
\IEEEauthorblockN{Samuel Todd Flanagan, Drupad K. Khublani, J.-F. Chamberland}
\IEEEauthorblockA{\textit{Dept. of ECE, Texas A\&M University} \\
College Station, USA \\
\{stflanagan, dkhublani, chmbrlnd\}@tamu.edu}
\and
\IEEEauthorblockN{Siddharth Agarwal, Ankit Vora}
\IEEEauthorblockA{
\textit{Ford Autonomous Vehicles LLC}\\
Dearborn, USA \\
\{sagarw20, avora3\}@ford.com}
}

\maketitle

\begin{abstract}
Fine localization is a crucial task for autonomous vehicles.
Although many algorithms have been explored in the literature for this specific task, the goal of getting accurate results from commodity sensors remains a challenge.
As autonomous vehicles make the transition from expensive prototypes to production items, the need for inexpensive, yet reliable solutions is increasing rapidly.
This article considers scenarios where images are captured with inexpensive cameras and localization takes place using pre-loaded fine maps of local roads as side information.
The techniques proposed herein extend schemes based on normalized mutual information by leveraging the likelihood of shades rather than exact sensor readings for localization in noisy environments.
This algorithmic enhancement, rooted in statistical signal processing, offers substantial gains in performance.
Numerical simulations are used to highlight the benefits of the proposed techniques in representative application scenarios.
Analysis of a Ford image set is performed to validate the core findings of this work.
\end{abstract}


\section{Introduction}

Localization is a cornerstone of vehicular autonomy.
The coarse location information afforded by exogenous systems based on satellites and cellular infrastructures is inadequate for autonomous vehicles, as current platforms rely on centimeter-scale accuracy.
Furthermore, wireless-based schemes are often unreliable in urban canyons where sky clearance may be narrow or nonexistent and structures create a scattering rich environment.
This situation has been widely recognized in the past, with engineers and designers turning to alternate means to acquire vehicle locations.
For instance, LIDAR technology is often found on prototype autonomous platforms.

As autonomous vehicles make the transition from research ventures to production items, design decisions are increasingly determined by anticipated cost.
At the same time, acquisition devices and the data they generate change from very precise observations to noisy measurements.
In this context, some of the assumptions that underlie localization algorithms, in terms of reliability and signal-to-noise ratio (SNR), may have to be revisited.
In this article, we explore the repercussions of performing localization based on images acquired by inexpensive cameras as opposed to alternate, more expensive devices.
More importantly, we identify algorithmic enhancements rooted in statistical signal processing that can substantially improve performance in noisy environments.

Characterizing and accounting for noise is a general topic of interest in signal processing. Techniques similar to those leveraged in this article have been used to remove colored noise from audio signals~\cite{kamath2002multi,majdak2011time} and nonuniform noise from images~\cite{chehdi1992new,hirakawa2005image}.
Specifically, Chehdi and Sabri's technique using maximum likelihood to define regions of uniform noise in images draws many similarities, if mostly in structure, with our work~\cite{chehdi1992new}.
Although some of the concepts are general, we focus our treatment on normalized mutual information (NMI) and its use in autonomous vehicles.

We initiate this discussion with a succinct overview of NMI, which is commonly used for medical image registration.
NMI is robust to intensity shifts and other intricacies common to changing outdoor conditions.
This property, along with past empirical successes, makes this quantity an appealing substratum for calibration and localization~\cite{pandey2015automatic,castorena2016autocalibration}.
Yet, current implementations of NMI-based localization algorithms tend to effectively give near absolute confidence to the pixel values provided by a camera.
Such approaches disregard the possibly noisy pixel acquisition process, as well as the fact that the quality of the information acquired by the camera may vary across pixels.
The realization that the information content in this setting is tightly coupled to the physics of image acquisition afford an algorithmic opportunity to improve performance.

Many past contributions on NMI and its applications pertain to medical image registration~\cite{pluim2003mutual}.
This task consists of aligning images with a computer model, or aligning features in an image with locations in physical space~\cite{hajnal2001medical}.
Collingnon, Maes, Viola, and Wells~\cite{collignon1995automated,maes1997multimodality,wells1996multi,viola1997alignment} spearheaded the use of mutual information (MI) for image registration.
Their work showed great results when MI was applied to rigid registration of multi-modality images, which made the technique popular.
Studholme et al. later proposed a normalized measure of mutual information (NMI) to make the technique more robust to outliers and changes in image overlap~\cite{studholme1999overlap}.
Herein, we adopt this formulation of NMI for localization.

The properties that make NMI effective for medical image registration also make it a viable technique for localization.
It is robust to light conditions, obstructions~\cite{dame2011mutual}, image overlap, and outliers~\cite{studholme1999overlap}.
NMI has begun to enter localization research because of these properties.
In 2014, Walcott and Eustice~\cite{wolcott2014visual} used a single monocular camera and NMI to perform localization within a 3D prior ground map, generated by 3D LIDAR data.
In their work, the authors generate synthetic views of the map in the perspective of the camera and use NMI to match them with live camera images.
Castorena and Agarwal~\cite{castorena2017ground} have also applied NMI to localization.
Their findings suggets that NMI slightly outperforms the state of the art and makes post-factory reflectivity calibration unnecessary, removing a major roadblock to production autonomous vehicles~\cite{castorena2017ground}.

Though NMI has shown some robustness to noise~\cite{dame2011mutual}, it does not explicitly account for SNR levels.
Moreover, the standard setting overlooks the unequal noise profile introduced by the physics of image acquisition for vehicular applications.
In this article, we propose a new, principled approach to NMI for vehicular localization.
Our methodology is rooted in both the physics of the problem and the statistical considerations that ensue.
We show that an enhanced NMI-based framework can improve localization, especially in poor conditions.

\section{{Problem Formulation}}
\label{section:ProblemFormulation}

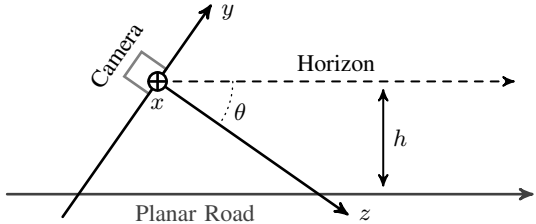
\begin{figure}[tbh]
\centerline{\input{FiguresNMI/CameraCoordinateSystem}}
\caption{This notional diagram offers an illustration of the camera, the road, and how they interact to yield images.
Variables $x$, $y$, and $z$ collectively define a coordinate system for the camera.}
\label{figure:CameraCoordinateSystem}
\end{figure}

We develop our ideas for monocular localization, although the concepts apply broadly.
As with many previously published techniques, the algorithm relies on a pre-built global map which is acquired over multiple passes and assumed noiseless.
The noise introduced as part of the image acquisition process is linked to the physics of the problem.
As such, we first seek to gain a better understanding of this process.
Our sensing device takes the form of a single pinhole camera with focal length $f$.
This camera is positioned at height $h$ above the road and inclined at angle $\theta$, as illustrated in Fig.~\ref{figure:CameraCoordinateSystem}.

\begin{figure}[tbh]
\centerline{\input{FiguresNMI/InternalCoordinateSystem}}
\caption{The external frame of reference (left) and the internal coordinate system (right) are chosen to simplify mathematical derivations.}
\label{figure:InternalCoordinateSystem}
\end{figure}
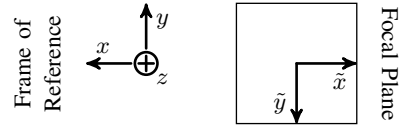

This setup yields two frames of reference.
One coordinate system is attached to the vehicle, the camera, and their surroundings.
We use $x$, $y$, and $z$ to denote its axes and set the origin at the pinhole of the camera, as shown in Fig.~\ref{figure:CameraCoordinateSystem}.
The second coordinate system is internal to the pinhole camera and dictated by the acquisition plane (e.g. CCD).
We employ variables $\tilde{x}$ and $\tilde{y}$ for this 2D internal coordinate system.
Under this characterization, the origin of the internal frame of reference is located at point $(0, 0, -f)$ in external coordinates.
The orientation of the internal coordinate system, shown in Fig.~\ref{figure:InternalCoordinateSystem}, is chosen such that $\tilde{x}$ and $\tilde{y}$ oppose $x$ and $y$.
This enables us to circumvent \emph{image inversion} in upcoming mathematical derivations.

\begin{figure}[tbh]
\centerline{\scalebox{0.8}{\input{FiguresNMI/GridRoad}} \scalebox{0.9}{\input{FiguresNMI/GridRoadImage}}}
\caption{The left diagram shows an arbitrary grid of equally sized squares on the road ahead of the autonomous vehicle.
The right diagram shows the grid pattern's image on the focal plane of the camera.
Squares with the same physical area do not project onto equally sized sections of the focal plane.}
\label{figure:GridRoad}
\end{figure}
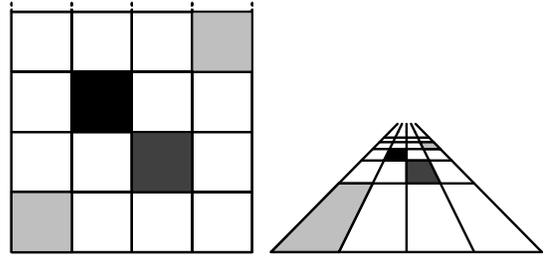

We mention that our analysis disregards potential barrel and pincushion distortion.
Thus, plane images captured by the camera simply undergo a perspective transformation, with
\begin{xalignat}{2} \label{equation:perspTransformation}
\tilde{x} &= {fx}/{z}  &
\tilde{y} &= {fy}/{z} .
\end{xalignat}
We employ a square grid tessellation of the road to highlight the effects of this transformation.
Figure~\ref{figure:GridRoad} displays a sample road segment in front of the camera along with its image.

\section{{Projection onto Focal Plane}}
\label{section:FocalPlane}

We will now review how a road section is projected onto the focal plane of the pinhole camera.
The flat road in Fig.~\ref{figure:CameraCoordinateSystem} has two degrees of freedom and can be captured using variables $x$ and $z$.
Then, $y$ becomes $z\tan\theta - h\sec\theta$.
Using this expression we rewrite $\tilde{y}$ in~\eqref{equation:perspTransformation} to be
\begin{equation}
\label{equation:yTildeFunctionOfZ}
\tilde{y} = f \tan \theta - ({f h}/{z}) \sec \theta .
\end{equation}
We also need to show $z$ in terms of $\tilde{y}$,
\begin{equation}
\label{equation:z}
z = \frac{f h}{f \sin \theta - \tilde{y} \cos \theta} .
\end{equation}
Using~\eqref{equation:perspTransformation} and~\eqref{equation:yTildeFunctionOfZ}, we can calculate the Jacobian of the perspective transformation, which is given by
\begin{equation}
\label{equation:Jacobian}
\Jm = \begin{bmatrix}
\frac{\partial \tilde{x}}{\partial x} & \frac{\partial \tilde{y}}{\partial x} \\
\frac{\partial \tilde{x}}{\partial z} & \frac{\partial \tilde{y}}{\partial z} \\
\end{bmatrix}
= \begin{bmatrix} \frac{f}{z} & 0 \\
- \frac{f x}{z^2} & - \frac{f h}{z^2} \sec \theta \end{bmatrix} .
\end{equation}
Recall the absolute value of the Jacobian determinant at a point on the road describes how the area near this point changes as it is projected onto the focal plane.
The determinant of~\eqref{equation:Jacobian} is $\operatorname{det}(\Jm) = - {f^2 h \sec \theta}/{z^3}$.
Let $\mathcal{R} = [x_{\mathrm{l}}, x_{\mathrm{u}}] \times [z_{\mathrm{l}}, z_{\mathrm{u}}]$ be a rectangular region ahead of the autonomous vehicle.
Using these properties, we deduce that the corresponding projected area on the focal plane of the camera is equal to
\begin{equation}
\label{equation:TildeA}
\tilde{\mathcal{A}} = f^2 h \sec \theta \frac{(x_{\mathrm{u}} - x_{\mathrm{l}})}{2} \left( \frac{1}{z_{\mathrm{l}}^2} - \frac{1}{z_{\mathrm{u}}^2} \right) .
\end{equation}
With this relationship defined, we next seek to characterize the profile of the noise introduced during image acquisition.

\section{{Noise Characterization}}
\label{section:NoiseCharacterization}

Every section of road captured by a camera undergoes a perspective transformation as its image gets projected onto the focal plane.
This transformation has significant implications regarding the noise introduced during image acquisition.
We define $q(x,z)$ as the function that captures the amplitude value at a point on the road at an instant in time.
Likewise, the amplitude value of this point projected on the focal plane is denoted by $\tilde{q} ( \tilde{x}, \tilde{y} )$.
Using~\eqref{equation:perspTransformation} and~\eqref{equation:z}, we can write
\begin{equation*}
\tilde{q} ( \tilde{x}, \tilde{y} )
= q \left( \frac{h \tilde{x}}{f \sin \theta - \tilde{y} \cos \theta}, \frac{f h}{f \sin \theta - \tilde{y} \cos \theta} \right) .
\end{equation*}
The signal acquired by the camera sensor as a function of position is given by $\tilde{q}( \tilde{x}, \tilde{y} ) + N ( \tilde{x}, \tilde{y} )$.
Noise process $N(\cdot, \cdot)$ is 2D white noise and has power spectral density $N_0$.
The signal acquired over camera region $\tilde{\mathcal{R}}$ becomes
\begin{equation*}
\begin{split}
&S =
\underbrace{\iint_{\tilde{\mathcal{R}}} \tilde{q}( \tilde{x}, \tilde{y} ) d\tilde{x} d\tilde{y}}_{\text{signal amplitude}}
+ \underbrace{\iint_{\tilde{\mathcal{R}}} N ( \tilde{x}, \tilde{y} ) d\tilde{x} d\tilde{y}}_{\text{additive noise}} .
\end{split}
\end{equation*}
Pixelated thermal noise, assuming the sensor is not operating in saturation, is effectively modeled as a zero-mean additive Gaussian random variable, with variance proportional to the footprint of the original region on the focal plane.
Parameter $N_0$, the 2D power spectral density, is constant throughout the sensing surface.
Below, we restrict our road model to gain further insight into the perspective transformation and its effects on image acquisition.

Consider a rectangular region of road $\mathcal{R} = [x_{\mathrm{l}}, x_{\mathrm{u}}] \times [z_{\mathrm{l}}, z_{\mathrm{u}}]$ that has a uniform amplitude, $q(x,z) = a$ for $(x,z) \in \mathcal{R}$.
We hope to find the quality of the signal corresponding to this region's projection on the focal plane.
The energy of the signal corresponding to $\mathcal{R}$ is
\begin{equation}
\label{equation:FocalPlaneEnergy}
\begin{split}
\left( \iint_{\tilde{\mathcal{R}}} \tilde{q}( \tilde{x}, \tilde{y} ) d\tilde{x} d\tilde{y} \right)^2
= \left( \iint_{\tilde{\mathcal{R}}} a d\tilde{x} d\tilde{y} \right)^2
= a^2 \tilde{\mathcal{A}}^2 .
\end{split}
\end{equation}
An expression for $\tilde{\mathcal{A}}$, the projected area of region $\mathcal{R}$ on the focal plane, appears in~\eqref{equation:TildeA}.
The noise introduced by the CCD is independent of $\tilde{q}( \tilde{x}, \tilde{y} )$ and has $\mu = 0$ and $\sigma^2 = \tilde{\mathcal{A}} N_0$, as discussed above.
The effective SNR of the observed signal corresponding to $\mathcal{R}$ is
\begin{equation}
\label{equation:SNR}
\operatorname{SNR}
= \frac{a^2 \tilde{\mathcal{A}}}{N_0}
= \frac{a^2}{N_0} \frac{f^2 h}{\cos \theta} \frac{(x_{\mathrm{u}}-x_{\mathrm{l}})}{2} \left( \frac{1}{z_{\mathrm{l}}^2} - \frac{1}{z_{\mathrm{u}}^2} \right) .
\end{equation}
Note the SNR decreases substantially as a function of $z$.
This relationship should be considered when matching a captured image with the global map.
The impact of this phenomenon will be most prevalent in poor conditions, with low SNR values.

\subsection{Road Tiles and the Likelihood Function}
\label{subsection:TileLikelihood}

To further illustrate this situation, we examine a road segment tessellated into squares, as shown in Fig.~\ref{figure:GridRoad}.
Each square has a uniform amplitude, which is randomly assigned independent of other squares.
When acquired by the camera, these regions will have distinct SNRs in accordance with~\eqref{equation:SNR}.

The partitioned road area captured by the camera is flattened into a vector, denoted by $\av$.
This yields an equivalent characterization we express as $\Vm = \av + \Nm$, where $\av = (a_1, \ldots, a_m)$ designates the amplitude vector and $\Nm = (N_1, \ldots, N_m)$ is the additive multivariate Gaussian noise with distribution $\mathcal{N} (\zerov, \Sigma)$.
The joint PDF of sample $\Vm$ is
\begin{equation*}
f_{\Vm} (\vv ; \av) = \frac{1}{\sqrt{(2 \pi)^{m} |\Sigma|}}
\exp \left( - \frac{1}{2} (\vv - \av)^{\mathrm{T}} \Sigma^{-1} (\vv - \av) \right) .
\end{equation*}
Consequently, the likelihood function for $\av$ is given by
\begin{align*}
\mathcal{L} (\av ; \vv)
& = \frac{1}{\sqrt{(2 \pi)^{m} |\Sigma|}}
\exp \left( - \frac{1}{2} (\av - \vv)^{\mathrm{T}} \Sigma^{-1} (\av - \vv) \right) .
\end{align*}
Since $\Sigma$ is a diagonal matrix, we can rewrite the likelihood function as
\begin{equation*}
\mathcal{L} (\av ; \vv)
= \prod_{k=1}^m \frac{1}{\sqrt{2 \pi} \sigma_k} \exp \left(- \frac{(a_k - v_k)^2}{2 \sigma_k^2} \right) .
\end{equation*}
In particular, the likelihood for one specific tile reduces to
\begin{equation} \label{equation:TileLikelihood}
\begin{split}
\mathcal{L} (a_k ; v_k)
&= \frac{1}{\sqrt{2 \pi} \sigma_k} \exp \left(- \frac{(a_k - v_k)^2}{2 \sigma_k^2} \right) .
\end{split}
\end{equation}
Using~\eqref{equation:SNR}, we get $\sigma_k^2 = {N_0}/{\tilde{\mathcal{A}}_k}$.
This structure can be employed to produce an estimate of the tile-pair distribution associated with the camera image and a candidate location.

\section{{Enhanced NMI Algorithm}}
\label{section:ENMI}

We now propose enhanced NMI (ENMI), an algorithmic improvement to the use of NMI in localization tasks.
This modified approach builds upon standard NMI techniques for image matching and leverages the tile likelihood structure derived in Section~\ref{subsection:TileLikelihood}.
We begin with a review of mutual information as it pertains to localization.

The goal of localization is to determine a vehicle's position by finding the section of global map that matches with the current captured image.
At a given time the vehicle captures an image and candidate sections of the global map are selected.
The most likely match is found to be the candidate section resulting in the highest NMI value with the captured image.

\begin{figure}[tbh]
\centerline{\input{FiguresNMI/JointDistributionElem}}
\caption{This diagram illustrates how one tile-pair contributes to the (empirical) joint distribution, before normalization.
Random variables $A$ and $B$ correspond to the amplitude values in the captured image and map section, respectively.
For simplicity, these values range from 0 to 3.
}
\label{figure:JointDistributionOriginal}
\end{figure}
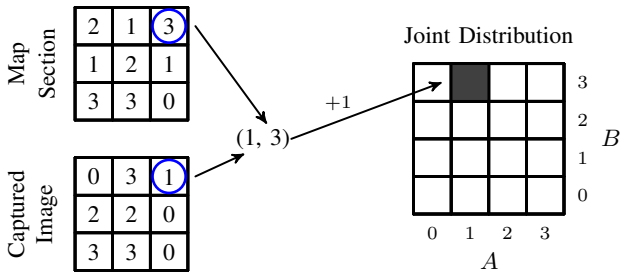

Computing the NMI of two tiled images is a straightforward process.
Given a captured image and map section, we couple their tile amplitude values based on position.
We subsequently bin the resulting pairs to generate a joint distribution of tile amplitude values, as depicted in Fig.~\ref{figure:JointDistributionOriginal}.
The joint distribution allows us to calculate the empirical NMI between the two images using the formulation proposed by Studholme et al.~\cite{studholme1999overlap}.
Specifically, let $(A,B)$ be a random vector drawn according to the empirical joint PMF described above, then
\begin{equation} \label{equation:NMI}
\begin{split}
\operatorname{NMI} [A, B]
&= \frac{H[A] + H[B]}{H[A, B]} ,
\end{split}
\end{equation}
where $H[A]$ and $H[B]$ are the (information) entropy of $A$ and $B$, respectively, and their joint entropy is $H[A, B]$.

The NMI equation in~\eqref{equation:NMI} is used in localization for matching captured images with candidate sections of the global map~\cite{castorena2017ground}.
For a given camera image, the NMI values it generates with the candidate sections are compared, with the section resulting in the largest NMI value being declared a match.
The autonomous vehicle determines its location based on the meta data of the selected map section.

\subsection{Likelihood-Based Joint Distribution}
\label{subsection:Likelihood-Based Joint Distribution}

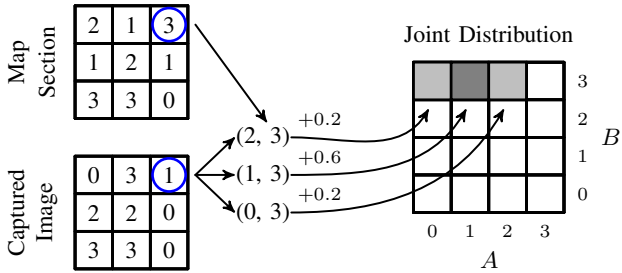
\begin{figure}[t!]
\centerline{\input{FiguresNMI/JointDistributionEnhancedElem}}
\caption{This diagram shows how the weight of one tile-pair is spread over multiple bins due to the uncertainty of the tile value in the captured image.
The distribution of weight depends on the effective SNR of the top right tile in the captured image, found using~\eqref{equation:SNR}.
}
\label{figure:JointDistributionEnhanced}
\end{figure}

We modify the procedure described above as to incorporate the uncertainty of noisy observations.
As before, the tile amplitude values in the captured image and the map section are paired based on position.
However, given that the captured image is noisy, we use the likelihood function defined in~\eqref{equation:TileLikelihood} to form a maximum a posteriori probability over possible tile amplitude values, rather than assigning a unit weight to the observation itself.
This action distributes the weight of each tile-pair over one or more bins depending on effective SNR as shown in Fig.~\ref{figure:JointDistributionEnhanced}.
We note that a similar operation could be performed on candidate map section values.
Yet, since we are assuming a noiseless global map, this latter action is unnecessary in the present context.
Once the joint distribution is computed using this approach, ENMI values are calculated with the standard NMI equation in~\eqref{equation:NMI}.
Just as before the candidate section of the global map resulting in the largest ENMI value is declared as a match, and its meta data is used to determine the vehicle's location.

\subsection{Simulated Performance}
\label{subsection:SimulatedPerformance}

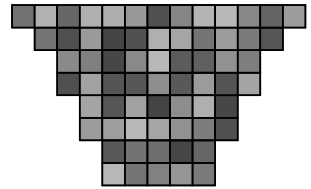
\begin{figure}[ht]
\begin{minipage}[c]{0.5\linewidth}
\centering
\begin{tabular}{l}
  \hline
  \textbf{Parameters} \tabularnewline
  \hline
  $f$ = 0.0367 cm \tabularnewline
  $\theta$ = 35.9020$^{\circ}$ \tabularnewline
  $h$ = 58.3095 cm \tabularnewline
  Side Length = 20 cm \tabularnewline
  Vertical View: 39.3$^{\circ}$ \tabularnewline
  Horizontal View: 70.5$^{\circ}$ \tabularnewline
  10,000 trials per point \tabularnewline
  \hline
\end{tabular}
\end{minipage}\hfill
\begin{minipage}[c]{0.5\linewidth}
\centering
  \scalebox{0.6}{\input{FiguresNMI/SimRoadImage}}
  \caption{This grid is an example of the rectified images used in simulations.}
  \label{figure:SimRoadImage}
\end{minipage}
\end{figure}

To illustrate the benefits of ENMI over NMI, we turn to numerical simulations.
Specifically, we evaluate the performance of ENMI and NMI for image matching by comparing their probabilities of error.
Using the following parameters we randomly generate gray-scale images of 66 squares with amplitude values assigned from a Gaussian distribution with $\mu = 128$ and $\sigma = 32$.
These images simulate captured sections of a grid road similar to Fig.~\ref{figure:GridRoad}.
After the images are rectified, the squares take on a v-shape as shown in Fig.~\ref{figure:SimRoadImage}.

We vary the power spectral density, $N_0$, to illustrate each technique's robustness to noise.
In each trial, we generate an image vector $\av$, noisy image vector $\Vm$, and candidate vector $\hat{\uv}$.
The noisy image vector $\Vm = \av + \Nm$ includes additive Gaussian noise with $\mu = 0$ and $\sigma^2 = {N_0}/{\tilde{\mathcal{A}}_k}$.
Each square in $\Vm$ has an effective SNR equal to~\eqref{equation:SNR}.
An error is recorded for NMI when $\operatorname{NMI} [\Vm,\hat{\uv}] \geq \operatorname{NMI} [\Vm,\av]$ and ENMI when $\operatorname{ENMI} [\Vm,\hat{\uv}] \geq \operatorname{ENMI} [\Vm,\av]$.
10,000 trials are performed for each value of $N_0$.

\begin{figure}[tbh]
\centerline{\scalebox{0.9}{\input{FiguresNMI/plot.tex}}}
\caption{This figure showcases the superior performance of ENMI over the standard NMI technique for noisy observations.}
\label{figure:simPerformance}
\end{figure}
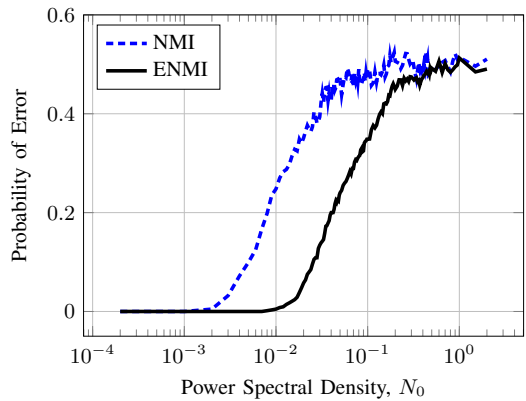

\section{{Ford Image Analysis}}

The motivation for ENMI relies on an unequal noise distribution in images captured by autonomous vehicles.
The noise profile developed in Section~\ref{section:NoiseCharacterization} follows the physics of our acquisition model, however, it is unclear whether this profile will present itself in real images.
We examine processed lidar images provided by Ford Autonomous Vehicles LLC to validate the noise profile.

\begin{figure}[tbph]
\centerline{\scalebox{1}{\includegraphics[scale=0.21]{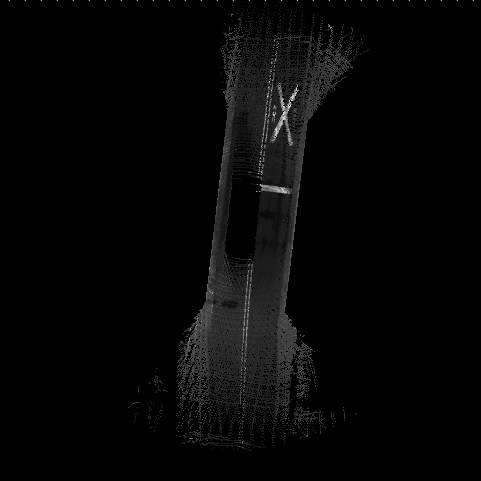}} \scalebox{1}{\includegraphics[scale=0.21]{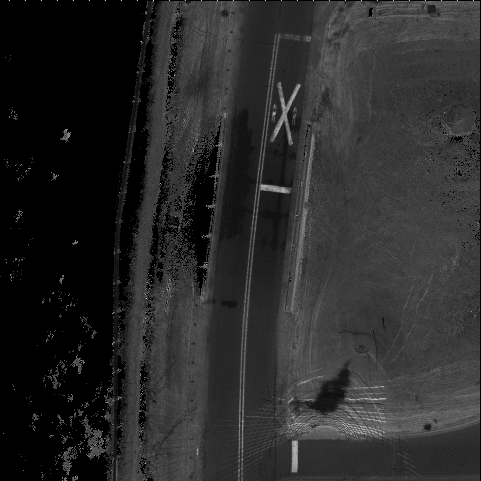}}}
\caption{The left image shows a current local image capture.
The right image shows the section of prior map that matches with the local image.}
\label{figure:LocalPriorImages}
\end{figure}

Figure~\ref{figure:LocalPriorImages} shows a matching pair of local and prior map images.
All images have the same orientation and are centered on the middle of the vehicle's rear axle.
We calculate the observed variance in pixel value at location $[i,j]$ using
\begin{equation*}
\begin{split}
\operatorname{Var[i,j]}
&= \frac{1}{n-1}\sum_{k=1}^{n} \left( a_{ij}^{(k)} - \tilde{a}_{ij}^{(k)} \right)^2,
\end{split}
\end{equation*}
where $a$ and $\tilde{a}$ are matching local and prior map images.

\begin{figure}[tbph]
\centering
\includegraphics[scale=0.53]{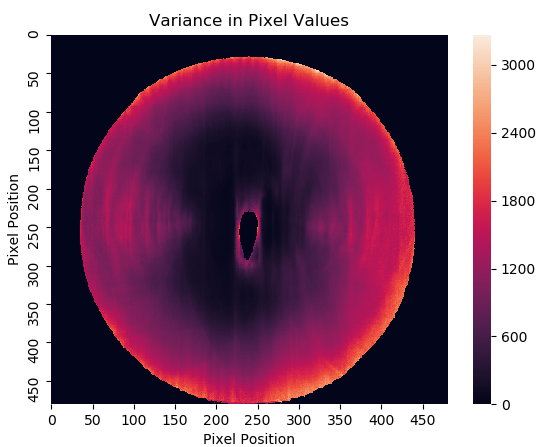}
\caption{This figure shows the observed variance in pixel values from the Ford image set.
The front of the vehicle is oriented facing the top of the image.
The black region around the outside of the mask is out of range.}
\label{figure:VarianceMask}
\end{figure}

We examine over 10,000 images, corrected for rotation, to produce the variance mask shown in Fig.~\ref{figure:VarianceMask}.
The variance increases radially but not exactly as described in Section~\ref{section:NoiseCharacterization}.
The disconnect should be expected because the formulations in this paper are derived based on an idealized model.
Still, the noise distribution is nonuniform.
This validates our assertion that equal confidence should not be given to all parts of an image when performing localization.
We plan to integrate such empirical masks in our future research efforts in this area.

\section{Discussion}

This article reviews the perspective transformation intrinsic to camera images and its effect on pixel reliability.
We use the properties of the transformation to characterize the noise profile it introduces.
We derive an enhanced NMI matching technique rooted in statistical signal processing and discuss how it applies to image matching.
This principled approach and the ensuing algorithm offer better performance for localization, especially in poor conditions.
This algorithmic improvement is timely.
As autonomous vehicles approach production, inexpensive, noisy sensors are increasingly used for localization.
In this sense, ENMI offers more reliable and robust image matching for future production applications.

\nocite{flanagan2019localization}

\bibliographystyle{IEEEbib}

\end{document}

%% file: FiguresNMI/CameraCoordinateSystem.tex
\begin{tikzpicture}[
  font=\small, >=stealth', line width=1.0pt, line cap=round
]

\def\mytheta{-35}

\draw[->, darkgray] (-2, -1.5) -- (5, -1.5) node[midway, below, xshift=-10mm] {Planar Road};

\draw[<->, line width=0.75] (3, -0.1) -- (3, -1.4) node[midway,right] {$h$};

\draw[dotted, line width=0.5] (1, 0) arc (0:-35:1);
\draw (1.1, -0.4) node[text=black] {$\theta$};

\draw[dashed, ->, line width=0.75] (0, 0) -- (4.75, 0) node[midway, above] {Horizon};

\draw[gray, rotate around={\mytheta:(0, 0)}] (-0.4, -0.2) rectangle (0, 0.2);
\node[rotate={90+\mytheta}] at (-0.55, 0.35) {Camera};



\draw[->] (0, 0) -- (\mytheta+90:1.25) node[anchor=west, yshift=-1mm] {$y$};
\draw[-] (0, 0) -- (\mytheta-90:2.2) node[anchor=east] {};
\draw[->] (0, 0) -- (\mytheta:3.1) node[anchor=west] {$z$};
\draw[fill=white] (0,0) circle (1.2mm) node[below, yshift=-1mm] {$x$};
\draw (-0.12,0) -- (0.12,0);
\draw (0,-0.12) -- (0,0.12);

\end{tikzpicture}

%% file: FiguresNMI/InternalCoordinateSystem.tex
\begin{tikzpicture}[
  font=\small, >=stealth', line width=1.0pt, line cap=round
]

\draw[line width=0.5] (-0.8, -0.8) rectangle (0.8, 0.8);
\draw (1, 0) node[anchor=south,rotate=-90] {Focal Plane};

\draw[->] (0, 0) -- (0, -0.8) node[anchor=south east] {$\tilde{y}$};
\draw[->] (0, 0) -- (0.8, 0) node[anchor=north east] {$\tilde{x}$};

\draw[->] (-2, 0.2) -- (-2, 0.8) node[anchor=north west] {$y$};
\draw[->] (-2.2, 0) -- (-2.8, 0) node[anchor=south west] {$x$};
\draw (-3, 0) node[anchor=south,rotate=90] {Reference};
\draw (-3.4, 0) node[anchor=south,rotate=90] {Frame of};

\draw (-2, 0) circle (0.15) node[anchor=north west] {$z$};
\draw (-1.9, 0) -- (-2.1, 0);
\draw (-2, -0.1) -- (-2, 0.1);

\end{tikzpicture}

%% file: FiguresNMI/GridRoad.tex
\begin{tikzpicture}[
  font=\small, >=stealth', line width=1.0pt, line cap=round
]

\foreach \z in {1,2,3,4,5} {
  \foreach \x in {-2,-1,0,1,2} {
    \draw (\x, 1) -- (\x, 5);
    \draw[dotted] (\x, 5) -- (\x, 5.15);
    \draw (-2, \z) -- (2, \z);
  }
}
\draw[fill=lightgray] (-1,1) rectangle (-2,2);
\draw[fill=darkgray] (1,2) rectangle (0,3);
\draw[fill=black] (0,3) rectangle (-1,4);
\draw[fill=lightgray] (2,4) rectangle (1,5);

\end{tikzpicture}

%% file: FiguresNMI/GridRoadImage.tex
\begin{tikzpicture}[
  font=\small, >=stealth', line width=1.0pt, line cap=round
]

\def\mytheta{-10}

\foreach \z in {1,2,3,4,5,6} {
  \draw ({-2/\z}, {tan(\mytheta) - (2*sec(\mytheta)/\z)}) -- ({2/\z}, {tan(\mytheta) - (2 * sec(\mytheta)/\z)});
}
\foreach \x in {-2,-1,0,1,2} {
  \draw (\x, {tan(\mytheta) - 2*sec(\mytheta)}) -- ({\x/8}, {tan(\mytheta) - (2 * sec(\mytheta)/8)});
  \draw ({\x/8}, {tan(\mytheta) - (2 * sec(\mytheta)/8)}) -- ({\x/15}, {tan(\mytheta) - (2 * sec(\mytheta)/15)});
}

\draw[fill=lightgray] (2/4,{tan(\mytheta) - (2 * sec(\mytheta)/4)})
-- (1/4,{tan(\mytheta) - (2 * sec(\mytheta)/4)})
-- (1/5,{tan(\mytheta) - (2 * sec(\mytheta)/5)})
-- (2/5,{tan(\mytheta) - (2 * sec(\mytheta)/5)})
-- (2/4,{tan(\mytheta) - (2 * sec(\mytheta)/4)});

\draw[fill=black] (-1/3,{tan(\mytheta) - (2 * sec(\mytheta)/3)})
-- (-1/4,{tan(\mytheta) - (2 * sec(\mytheta)/4)})
-- (0,{tan(\mytheta) - (2 * sec(\mytheta)/4)})
|- (-1/3,{tan(\mytheta) - (2 * sec(\mytheta)/3});

\draw[fill=darkgray] (1/2,{tan(\mytheta) - (2 * sec(\mytheta)/2)})
-- (1/3,{tan(\mytheta) - (2 * sec(\mytheta)/3)})
-- (0,{tan(\mytheta) - (2 * sec(\mytheta)/3)})
|- (1/2,{tan(\mytheta) - (2 * sec(\mytheta)/2)});

\draw[fill=lightgray] (-2,{tan(\mytheta) - (2 * sec(\mytheta))})
-- (-1,{tan(\mytheta) - (2 * sec(\mytheta)/2)})
-- (-1/2,{tan(\mytheta) - (2 * sec(\mytheta)/2)})
-- (-1,{tan(\mytheta) - (2 * sec(\mytheta))})
-- (-2,{tan(\mytheta) - (2 * sec(\mytheta))});

\end{tikzpicture}

%% file: FiguresNMI/JointDistributionElem.tex
\begin{tikzpicture}[
  font=\small, >=stealth', line width=1.0pt, line cap=round
]

\foreach \y in {0,1,2,3} {
  \foreach \x in {0,1,2,3} {
    \draw (0.5*\x, 0.25) -- (0.5*\x, 1.75);
    \draw (0, 0.25+0.5*\y) -- (1.5, 0.25+0.5*\y);
  }
}
\node[rotate=90] at (-0.75,1) {Captured};
\node[rotate=90] at (-0.375,1) {Image};
\node (kvalue) at (1.25, 1.5) {1};
\node at (0.75, 1.5) {3};
\node at (0.25, 1.5) {0};
\node at (0.25, 1) {2};
\node at (0.75, 1) {2};
\node at (1.25, 1) {0};
\node at (0.25, 0.5) {3};
\node at (0.75, 0.5) {3};
\node at (1.25, 0.5) {0};

\foreach \y in {0,1,2,3} {
  \foreach \x in {0,1,2,3} {
    \draw (0.5*\x, 2.25) -- (0.5*\x, 3.75);
    \draw (0, 2.25+0.5*\y) -- (1.5, 2.25+0.5*\y);
  }
}
\node[rotate=90] at (-0.75,3) {Map};
\node[rotate=90] at (-0.375,3) {Section};
\node (lvalue) at (1.25, 3.5) {3};
\node at (0.75, 3.5) {1};
\node at (0.25, 3.5) {2};
\node at (0.25, 3) {1};
\node at (0.75, 3) {2};
\node at (1.25, 3) {1};
\node at (0.25, 2.5) {3};
\node at (0.75, 2.5) {3};
\node at (1.25, 2.5) {0};

\node (pair) at (2.5, 2) {(1, 3)};
\draw[->, line width=0.75] (1.6,1.5) -- (2.25,1.80);
\draw[->, line width=0.75] (1.6,3.5) -- (2.55,2.20);

\draw[blue] (1.25, 1.5) circle (0.22);
\draw[blue] (1.25, 3.5) circle (0.22);

\foreach \y in {0,1,2,3,4} {
  \foreach \x in {0,1,2,3,4} {
    \draw (4.5+0.5*\x, 1.00) -- (4.5+0.5*\x, 3.00);
    \draw (4.5, 1+0.5*\y) -- (6.5, 1+0.5*\y);
  }
}
\node (JD) at (5.5, 3.375) {Joint Distribution};

\foreach \p in {0,1,2,3} {
  \node at (6.75, 1.25+0.5*\p) {\scriptsize $\p$};
  \node at (4.75+0.5*\p, 0.75) {\scriptsize $\p$};
}
\draw (5.25, 0.375) node[anchor=west] {$A$};
\draw (7.125, 2.25) node[anchor=north] {$B$};

\draw[fill=darkgray] (5.0, 2.5) rectangle (5.5, 3);

\draw[->, line width=0.75] (2.875, 2) -- (4.875, 2.75);
\node[anchor=south] at (3.5,2.25) {\scriptsize $+1$};
\end{tikzpicture}

%% file: FiguresNMI/JointDistributionEnhancedElem.tex
\begin{tikzpicture}[
  font=\small, >=stealth', line width=1.0pt, line cap=round
]

\foreach \y in {0,1,2,3} {
  \foreach \x in {0,1,2,3} {
    \draw (0.5*\x, 0.25) -- (0.5*\x, 1.75);
    \draw (0, 0.25+0.5*\y) -- (1.5, 0.25+0.5*\y);
  }
}
\node[rotate=90] at (-0.75,1) {Captured};
\node[rotate=90] at (-0.375,1) {Image};
\node (kvalue) at (1.25, 1.5) {1};
\node at (0.75, 1.5) {3};
\node at (0.25, 1.5) {0};
\node at (0.25, 1) {2};
\node at (0.75, 1) {2};
\node at (1.25, 1) {0};
\node at (0.25, 0.5) {3};
\node at (0.75, 0.5) {3};
\node at (1.25, 0.5) {0};

\foreach \y in {0,1,2,3} {
  \foreach \x in {0,1,2,3} {
    \draw (0.5*\x, 2.25) -- (0.5*\x, 3.75);
    \draw (0, 2.25+0.5*\y) -- (1.5, 2.25+0.5*\y);
  }
}
\node[rotate=90] at (-0.75,3) {Map};
\node[rotate=90] at (-0.375,3) {Section};
\node (lvalue) at (1.25, 3.5) {3};
\node at (0.75, 3.5) {1};
\node at (0.25, 3.5) {2};
\node at (0.25, 3) {1};
\node at (0.75, 3) {2};
\node at (1.25, 3) {1};
\node at (0.25, 2.5) {3};
\node at (0.75, 2.5) {3};
\node at (1.25, 2.5) {0};

\node (pair2) at (2.5, 2) {(2, 3)};
\node (pair1) at (2.5, 1.5) {(1, 3)};
\node (pair0) at (2.5, 1.0) {(0, 3)};
\draw[->, line width=0.75] (1.6,3.5) -- (2.55,2.20);

\draw[->, line width=0.75] (1.6,1.5) -- (2.125,2.0);
\draw[->, line width=0.75] (1.6,1.5) -- (2.125,1.5);
\draw[->, line width=0.75] (1.6,1.5) -- (2.125,1.0);

\draw[blue] (1.25, 1.5) circle (0.22);
\draw[blue] (1.25, 3.5) circle (0.22);

\foreach \y in {0,1,2,3,4} {
  \foreach \x in {0,1,2,3,4} {
    \draw (4.5+0.5*\x, 1.00) -- (4.5+0.5*\x, 3.00);
    \draw (4.5, 1+0.5*\y) -- (6.5, 1+0.5*\y);
  }
}
\node (JD) at (5.5, 3.375) {Joint Distribution};

\foreach \p in {0,1,2,3} {
  \node at (6.75, 1.25+0.5*\p) {\scriptsize $\p$};
  \node at (4.75+0.5*\p, 0.75) {\scriptsize $\p$};
}
\draw (5.25, 0.375) node[anchor=west] {$A$};
\draw (7.125, 2.25) node[anchor=north] {$B$};

\draw[fill=lightgray] (4.5, 2.5) rectangle (5, 3);
\draw[fill=gray] (5, 2.5) rectangle (5.5, 3);
\draw[fill=lightgray] (5.5, 2.5) rectangle (6, 3);

\draw[->, line width=0.75] (2.875, 2) to [out=0,in=-120] (4.7, 2.375);
\draw[->, line width=0.75] (2.875, 1.5) to [out=0,in=-120] (5.2, 2.375);
\draw[->, line width=0.75] (2.875, 1) to [out=0,in=-120] (5.7, 2.375);
\node[anchor=south] at (3.25, 2) {\scriptsize $+0.2$};
\node[anchor=south] at (3.25, 1.5) {\scriptsize $+0.6$};
\node[anchor=south] at (3.25, 1) {\scriptsize $+0.2$};

\end{tikzpicture}

%% file: FiguresNMI/SimRoadImage.tex
\begin{tikzpicture}[
  font=\small, >=stealth', line width=1.0pt, line cap=round
]


\foreach \i in {0, 0.5} {
    \foreach \j in {-1.25, -0.75, -0.25, 0.25, 0.75} {
        \pgfmathparse{50*rnd + 25}\edef\temp{\pgfmathresult}
        \draw[fill=black!\temp!white] (\j, \i) rectangle (\j + 0.5, \i + 0.5);
    }
}
\foreach \i in {1, 1.5} {
    \foreach \j in {-1.75, -1.25, -0.75, -0.25, 0.25, 0.75, 1.25} {
        \pgfmathparse{50*rnd + 25}\edef\temp{\pgfmathresult}
        \draw[fill=black!\temp!white] (\j, \i) rectangle (\j + 0.5, \i + 0.5);
    }
}
\foreach \i in {2, 2.5} {
    \foreach \j in {-2.25, -1.75, -1.25, -0.75, -0.25, 0.25, 0.75, 1.25, 1.75} {
        \pgfmathparse{50*rnd + 25}\edef\temp{\pgfmathresult}
        \draw[fill=black!\temp!white] (\j, \i) rectangle (\j + 0.5, \i + 0.5);
    }
}
\foreach \i in {3} {
    \foreach \j in {-2.75, -2.25, -1.75, -1.25, -0.75, -0.25, 0.25, 0.75, 1.25, 1.75, 2.25} {
        \pgfmathparse{50*rnd + 25}\edef\temp{\pgfmathresult}
        \draw[fill=black!\temp!white] (\j, \i) rectangle (\j + 0.5, \i + 0.5);
    }
}
\foreach \i in {3.5} {
    \foreach \j in {-3.25, -2.75, -2.25, -1.75, -1.25, -0.75, -0.25, 0.25, 0.75, 1.25, 1.75, 2.25, 2.75} {
        \pgfmathparse{50*rnd + 25}\edef\temp{\pgfmathresult}
        \draw[fill=black!\temp!white] (\j, \i) rectangle (\j + 0.5, \i + 0.5);
    }
}

\foreach \x in {-1.25,-0.75,-0.25,0.25,0.75,1.25} {
  \draw (\x, 0) -- (\x, 4);
}

\foreach \z in {0,0.5,1,1.5,2,2.5,3,3.5,4} {
  \draw (-1.25, \z) -- (1.25, \z);
}

\draw (-1.75, 1) -- (-1.75, 4);
\draw (1.75, 1) -- (1.75, 4);
\draw (-2.25, 2) -- (-2.25, 4);
\draw (2.25, 2) -- (2.25, 4);
\draw (-2.75, 3) -- (-2.75, 4);
\draw (2.75, 3) -- (2.75, 4);
\draw (-3.25, 3.5) -- (-3.25, 4);
\draw (3.25, 3.5) -- (3.25, 4);

\draw (-1.75, 1) -- (1.75, 1);
\draw (-1.75, 1.5) -- (1.75, 1.5);
\draw (-2.25, 2) -- (2.25, 2);
\draw (-2.25, 2.5) -- (2.25, 2.5);
\draw (-2.75, 3) -- (2.75, 3);
\draw (-3.25, 3.5) -- (3.25, 3.5);
\draw (-3.25, 4) -- (3.25, 4);

\end{tikzpicture}

%% file: FiguresNMI/plot.tex
\begin{tikzpicture}

\begin{axis}[
font=\small,
scale only axis,
width=6.5cm,
height=4.75cm,
xmode=log,
ymin=-0.05, ymax=0.6,
xlabel={Power Spectral Density, $N_0$},
ylabel={Probability of Error},
xmajorgrids,
ymajorgrids,
zmajorgrids,
legend entries={NMI, ENMI},
legend style={nodes=right},
legend pos=north west]

\addplot [color=blue, densely dashed, line width=1.5pt]
coordinates{
(0.0002,0.0)
(0.0004,0.0)
(0.0006,0.0)
(0.0008,0.0)
(0.001,0.0)
(0.002,0.0048)
(0.003,0.032)
(0.004,0.072)
(0.005,0.0992)
(0.006,0.1248)
(0.007,0.1668)
(0.008,0.1984)
(0.009,0.2354)
(0.01,0.2479)
(0.011,0.2703)
(0.012,0.2827)
(0.013,0.2874)
(0.014,0.2974)
(0.015,0.3152)
(0.016,0.328)
(0.017,0.3248)
(0.018,0.3485)
(0.019,0.3405)
(0.02,0.351)
(0.021,0.3646)
(0.022,0.3765)
(0.023,0.3694)
(0.024,0.364)
(0.025,0.3736)
(0.026,0.4039)
(0.027,0.3964)
(0.028,0.3808)
(0.029,0.3846)
(0.03,0.3935)
(0.031,0.4221)
(0.032,0.441)
(0.033,0.4232)
(0.034,0.4433)
(0.035,0.4529)
(0.036,0.4535)
(0.037,0.4268)
(0.038,0.4426)
(0.039,0.427)
(0.04,0.4251)
(0.041,0.4538)
(0.042,0.4376)
(0.043,0.4631)
(0.044,0.4485)
(0.045,0.4531)
(0.046,0.4688)
(0.047,0.4697)
(0.048,0.4614)
(0.049,0.4691)
(0.05,0.4691)
(0.052,0.4387)
(0.054,0.4589)
(0.056,0.4756)
(0.058,0.473)
(0.06,0.4922)
(0.062,0.4845)
(0.064,0.468)
(0.066,0.4663)
(0.068,0.4806)
(0.07,0.4573)
(0.072,0.4824)
(0.074,0.4679)
(0.076,0.4788)
(0.078,0.4628)
(0.08,0.4899)
(0.082,0.4655)
(0.084,0.4589)
(0.086,0.4612)
(0.088,0.4672)
(0.09,0.4705)
(0.092,0.4833)
(0.094,0.4656)
(0.096,0.4831)
(0.098,0.4713)
(0.1,0.4907)
(0.105,0.4873)
(0.11,0.4574)
(0.115,0.4625)
(0.12,0.4846)
(0.125,0.4638)
(0.13,0.4813)
(0.135,0.4815)
(0.14,0.4787)
(0.145,0.4856)
(0.15,0.4983)
(0.155,0.4745)
(0.16,0.4807)
(0.165,0.495)
(0.17,0.498)
(0.175,0.5183)
(0.18,0.5037)
(0.185,0.5137)
(0.19,0.5168)
(0.195,0.5239)
(0.2,0.5062)
(0.21,0.5041)
(0.22,0.5019)
(0.23,0.4913)
(0.24,0.4977)
(0.25,0.5105)
(0.26,0.5219)
(0.27,0.5214)
(0.28,0.5103)
(0.29,0.5011)
(0.3,0.5055)
(0.31,0.5088)
(0.32,0.4942)
(0.33,0.481)
(0.34,0.4702)
(0.35,0.4864)
(0.36,0.5037)
(0.37,0.4916)
(0.38,0.5012)
(0.39,0.4847)
(0.4,0.4881)
(0.41,0.5007)
(0.42,0.4921)
(0.43,0.5247)
(0.44,0.5012)
(0.45,0.5109)
(0.46,0.4818)
(0.47,0.4848)
(0.48,0.4851)
(0.49,0.4883)
(0.5,0.4883)
(0.55,0.4895)
(0.6,0.4976)
(0.65,0.5038)
(0.7,0.5078)
(0.75,0.4959)
(0.8,0.485)
(0.85,0.5069)
(0.9,0.5214)
(0.95,0.5106)
(1.0,0.5139)
(1.5,0.4962)
(2.0,0.5105)
};

\addplot [color=black, solid, line width=1.5pt]
coordinates{
(0.0002,0.0)
(0.0004,0.0)
(0.0006,0.0)
(0.0008,0.0)
(0.001,0.0)
(0.002,0.0)
(0.003,0.0)
(0.004,0.0)
(0.005,0.0)
(0.006,0.0)
(0.007,0.0)
(0.008,0.0016)
(0.009,0.0032)
(0.01,0.0048)
(0.011,0.008)
(0.012,0.0095)
(0.013,0.0143)
(0.014,0.0175)
(0.015,0.0208)
(0.016,0.024)
(0.017,0.0288)
(0.018,0.0384)
(0.019,0.0479)
(0.02,0.0576)
(0.021,0.0639)
(0.022,0.0752)
(0.023,0.08)
(0.024,0.0848)
(0.025,0.096)
(0.026,0.1055)
(0.027,0.1088)
(0.028,0.1088)
(0.029,0.1276)
(0.03,0.1341)
(0.031,0.1388)
(0.032,0.142)
(0.033,0.1455)
(0.034,0.1581)
(0.035,0.1726)
(0.036,0.174)
(0.037,0.179)
(0.038,0.1837)
(0.039,0.195)
(0.04,0.2)
(0.041,0.2)
(0.042,0.2001)
(0.043,0.2063)
(0.044,0.2175)
(0.045,0.2238)
(0.046,0.2206)
(0.047,0.2256)
(0.048,0.2239)
(0.049,0.2353)
(0.05,0.2401)
(0.052,0.245)
(0.054,0.255)
(0.056,0.2597)
(0.058,0.263)
(0.06,0.2693)
(0.062,0.276)
(0.064,0.2832)
(0.066,0.2783)
(0.068,0.2896)
(0.07,0.2895)
(0.072,0.3023)
(0.074,0.3072)
(0.076,0.3056)
(0.078,0.3103)
(0.08,0.3136)
(0.082,0.3157)
(0.084,0.3269)
(0.086,0.3302)
(0.088,0.3232)
(0.09,0.3248)
(0.092,0.3408)
(0.094,0.3439)
(0.096,0.3453)
(0.098,0.3485)
(0.1,0.3486)
(0.105,0.3489)
(0.11,0.3679)
(0.115,0.3761)
(0.12,0.3715)
(0.125,0.3767)
(0.13,0.3895)
(0.135,0.3989)
(0.14,0.4039)
(0.145,0.4004)
(0.15,0.4201)
(0.155,0.4215)
(0.16,0.4201)
(0.165,0.4249)
(0.17,0.4375)
(0.175,0.4364)
(0.18,0.4477)
(0.185,0.4557)
(0.19,0.4606)
(0.195,0.4555)
(0.2,0.4494)
(0.21,0.4558)
(0.22,0.4687)
(0.23,0.4685)
(0.24,0.4573)
(0.25,0.4719)
(0.26,0.4666)
(0.27,0.4704)
(0.28,0.4654)
(0.29,0.475)
(0.3,0.4745)
(0.31,0.4714)
(0.32,0.4683)
(0.33,0.471)
(0.34,0.4727)
(0.35,0.4649)
(0.36,0.4673)
(0.37,0.4579)
(0.38,0.465)
(0.39,0.4709)
(0.4,0.4728)
(0.41,0.4778)
(0.42,0.4791)
(0.43,0.4823)
(0.44,0.4857)
(0.45,0.4825)
(0.46,0.4842)
(0.47,0.4793)
(0.48,0.4919)
(0.49,0.4872)
(0.5,0.4921)
(0.55,0.486)
(0.6,0.5063)
(0.65,0.4836)
(0.7,0.5015)
(0.75,0.4889)
(0.8,0.4839)
(0.85,0.4892)
(0.9,0.495)
(0.95,0.4985)
(1.0,0.5128)
(1.5,0.4847)
(2.0,0.4904)
};

\end{axis}
\end{tikzpicture}